\documentclass[sigconf]{acmart}

\AtBeginDocument{%
  }

\setcopyright{acmlicensed}
\copyrightyear{2018}
\acmYear{2018}
\acmDOI{XXXXXXX.XXXXXXX}
\acmConference[JAWs 2026]{Make sure to enter the correct
  conference title from your rights confirmation email}{April 12--18,
  2026}{Rio De Janeiro, Brazil}
\acmBooktitle{JAWs '26: ACM 1st Journal Ahead Workshop}
\acmISBN{978-1-4503-XXXX-X/2018/06}

\usepackage{subcaption}
\usepackage{tabularx}
\usepackage{enumitem}
\usepackage{pifont}
\usepackage{pgfplots}
\usepackage[dvipsnames]{xcolor} 
\usepackage[normalem]{ulem}
\usepackage{array}
\usepackage{lipsum}
\usepackage{hyperref}
\usepackage{tabularray} 
\usepackage{graphicx}
\usepackage{caption}
\usepackage{tikz}
\usepackage{ragged2e}
\usepackage[most]{tcolorbox}
\usetikzlibrary{patterns}
\usepackage{fontawesome}

\UseTblrLibrary{booktabs} 
\usetikzlibrary{calc}

\hypersetup{
    colorlinks=true,
    linkcolor=blue,
    filecolor=magenta,      
    urlcolor=cyan,
    pdftitle={Overleaf Example},
    pdfpagemode=FullScreen,
    }
\usepackage{amsmath}

\definecolor{mygrayzero}{gray}{0.9}
\definecolor{mygrayone}{gray}{0.8}
\definecolor{mygraytwo}{gray}{0.7}
\definecolor{mygraythree}{gray}{0.6}

\definecolor{myblue}{RGB}{230, 240, 255} 
\definecolor{myborder}{RGB}{100, 100, 255} 

\definecolor{Mycolor}{HTML}{166666}
\definecolor{revision}{HTML}{D10000}

\begin{document}

\title{From Generic to Personalized: Exploring Persona-Aware Code Review Explanations}






\author{Shamse Tasnim Cynthia\textsuperscript{$\diamondsuit$}, Ratnadira Widyasari\textsuperscript{$\clubsuit$}, Banani Roy\textsuperscript{$\diamondsuit$}, Italo Santos\textsuperscript{$\spadesuit$}, and David Lo\textsuperscript{$\clubsuit$}}

\affiliation{%
  \institution{\textsuperscript{$\diamondsuit$}University of Saskatchewan, Canada \textsuperscript{$\clubsuit$}Singapore Management University, Singapore}\country{}
}
\affiliation{%
  \institution{\textsuperscript{$\spadesuit$}University of Hawai'i, USA}\country{}
}

\affiliation{%
  \institution{\{shamse.cynthia, banani.roy\}@usask.ca, \{ratnadiraw.2020, davidlo\}@smu.edu.sg, isantos3@hawaii.edu}\country{}
}


\renewcommand{\shortauthors}{Shamse Tasnim Cynthia, Ratnadira Widyasari, Banani Roy, Italo Santos, and David Lo}

\begin{abstract}
  Code review is essential for ensuring software quality and supporting collaboration, yet prior work shows that developers can interpret code review comments differently. These differences can hinder effective communication, particularly in collaborative settings. 
  To address this challenge, we explore the potential of personified code review explanations. We report initial findings from an ongoing mixed-methods user study in which developers evaluated persona-aligned review comments across multiple code snippets. Our results suggest that preferences for explanation styles vary across problem-solving styles, experience levels, and roles. 
  Across problem-solving style profiles, developers valued explanatory depth, learning support, practical suggestions, and risk awareness over conciseness, highlighting the need to balance personalization with clarity and trust.
  Based on these findings, we outline a vision for inclusive, human-centered AI-assisted code review systems that adapt feedback to developers’ problem-solving preferences.
\end{abstract}





\maketitle

\section{Introduction} \label{sec:introduction}
Code review is a widely used technique for systematically examining code changes, aiming to increase software quality~\cite{mcintosh2014impact}. Code reviews provide several benefits for the project, including finding bugs, knowledge transfer, and assurance of adherence to project guidelines and coding style~\cite{morales2015code, bavota2015four}. 
An effective code review depends on active involvement and cooperation between reviewers and code authors. Prior research has identified several key characteristics of high-quality review comments, showing that explanatory feedback and a constructive tone promote understanding, whereas harsh language, superficial critiques, and impractical suggestions hinder knowledge transfer~\cite{sarma2024effective, turzo2024makes, chen2025understanding}. Moreover, review comments are perceived as more helpful when they help build positive relationships among developers~\cite{bosu2016process}.
However, recent findings by Widyasari et al.~\cite{widyasari2025explaining} indicate that the interpretation of code review explanations can vary substantially across developers. What appears obvious to one developer may not be equally clear to another, which might reflect the underlying cognitive diversity in how developers reason about, process, and act on feedback~\cite{hyrynsalmi2025making}. 
This oversight results in a critical research gap since current automated and manual code review systems rarely adapt explanations to individual reviewers' problem-solving styles (e.g., differences in information processing, learning orientation, and motivation~\cite{burnett2016gendermag, padala2020gender}) or experience level (novice vs. expert). As a result, explanations may be misaligned, misunderstood, or even ignored, which leads to increased back-and-forth, slower reviews, and abandoned pull requests \cite{ khatoonabadi2023wasted, vitale2025personalized}.
Therefore, a systematic investigation is required to find out the effectiveness of personalized, need-driven code review comments that better support knowledge sharing among developers.

Recent advances in Large Language Models (LLMs) present a promising opportunity to bridge this gap. Prompt-engineered LLMs can generate explanations with varying tones, formats, and levels of detail~\cite{santos2025great, brachman2025towards}. When combined with problem-solving profiling approaches such as the GenderMag framework~\cite{burnett2016finding}, these models can be guided to produce persona-aligned code review explanations tailored to individual differences in problem-solving styles~\cite{anderson2024measuring, choudhuri2025needs}.

Our study aims to explore how persona-aligned code review explanations influence developers’ trust, comprehension, and decision-making. Unlike most prior studies that treat developers uniformly and overlook cognitive diversity~\cite{alami2025engagement, chen2025understanding}, we build on an initial mixed-methods user study to examine how developers with different problem-solving styles perceive the usefulness and value of GenderMag persona-aligned code review comments. Our goal is to inform the design of inclusive and human-centered AI-assisted code review systems. In this paper, we outline our research goals, present preliminary results from our user study, and discuss the overall impact.

\vspace{-0.5cm}
\section{Motivational Example} \label{sec:motivational-example}

\begin{figure}[t]
    \centering
    \begin{subfigure}[t]{0.85\linewidth}
        \centering
        \includegraphics[width=\linewidth]{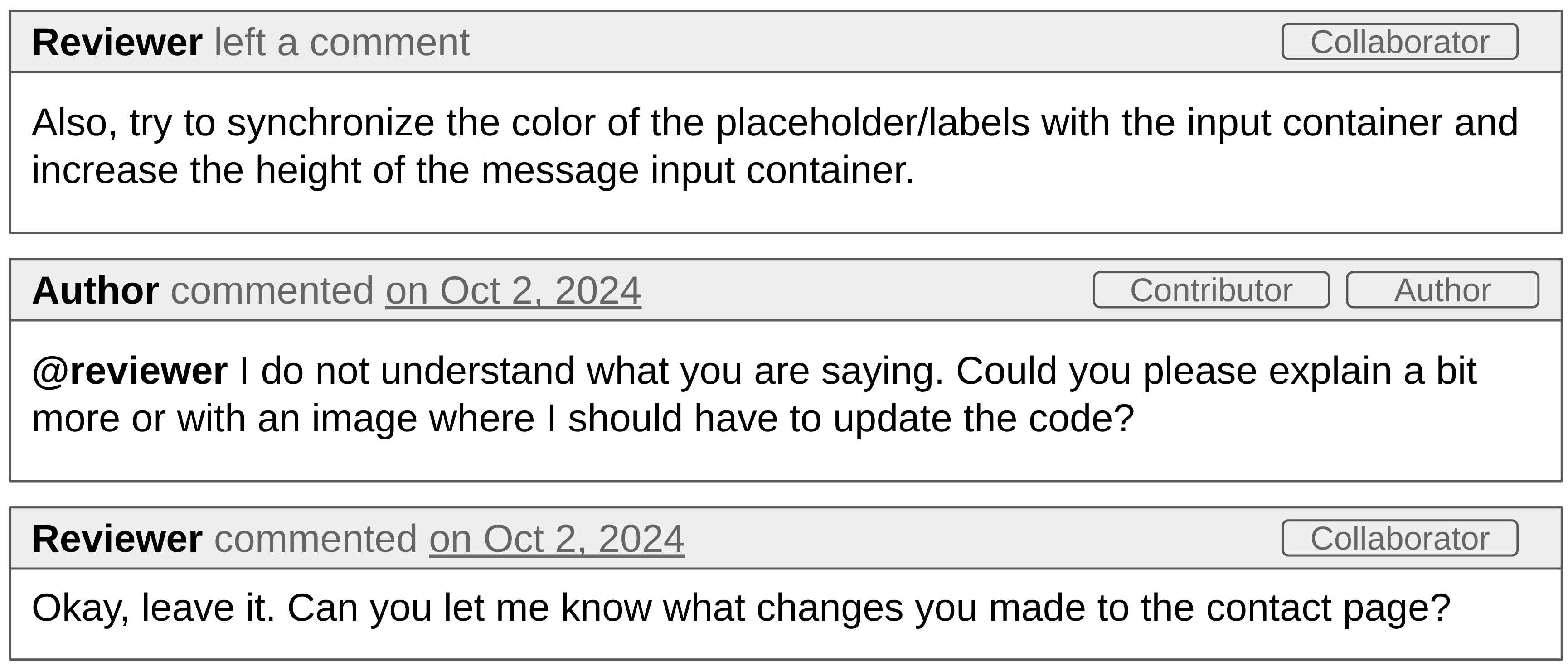}
        \caption{Example 1}
        \label{fig:motiv1}
    \end{subfigure}
    \vspace{0.6em}
    \begin{subfigure}[t]{0.85\linewidth}
        \centering
        \includegraphics[width=\linewidth]{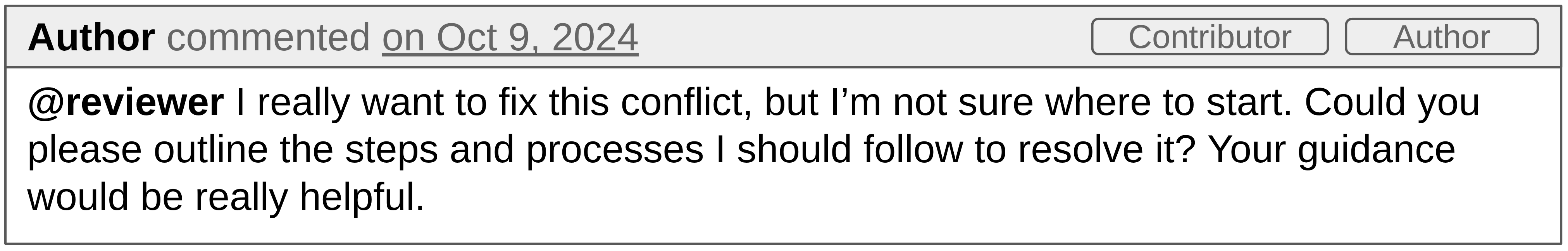}
        \caption{Example 2}
        \vspace{-0.5cm}
        \label{fig:motiv2}
    \end{subfigure}
    \caption{Example of code review comments misaligned with developers' problem-solving styles \cite{motiv1, motiv2}.}
    \label{fig:motivational-examples}
\vspace{-0.7cm}
\end{figure}

Fig.~\ref{fig:motiv1} illustrates a code review comment from a GitHub pull request~\cite{motiv1} that highlights challenges arising from underspecified and cognitively misaligned explanations. The reviewer suggests a design improvement but provides limited context regarding the affected components, the issue location, or the rationale for the suggestion. As a result, the contributor is unable to interpret the feedback and requests clarification. The discussion subsequently shifts away from the original comment without resolving the initial concern, demonstrating how misaligned explanations can hinder effective communication and knowledge transfer in code review.

Fig.~\ref{fig:motiv2}~\cite{motiv2} represents another example where a contributor requests guidance for resolving a complex issue, reflecting a need for a structured, procedural explanation. The reviewer instead resolves the issue directly and provides a brief, high-level explanation. 
Although the interaction concludes successfully, it exposes a misalignment between the explanations offered and the developers’ actual needs. These examples motivate the need for adaptive code review explanations tailored to developers’ problem-solving preferences and information requirements.

\section{Research Questions and Methodology} \label{sec:RQs-methodology}
This study aims to establish an empirically grounded understanding of how cognitive diversity shapes developers' interactions with code review explanations and how this understanding can inform more inclusive code review practices. Rather than assuming that a single explanation type is universally effective, we aim to examine how developers with different problem-solving styles perceive and engage with code review feedback. Thus, we seek to answer the following research questions (RQs):

    \textbf{RQ\textsubscript{1}.}~\textbf{How do developers' preferences for code review explanations differ according to their problem-solving style, experience, and roles in the development?}
    Developers vary in how they process information and respond to feedback, suggesting that code review explanations may not be uniformly effective \cite{widyasari2025explaining, anderson2025llm}. Problem-solving style, development experience, and role in the review process can each influence the type of explanation developers find most useful. Thus, in this RQ, we aim to examine whether and how these individual differences translate into distinct preferences for code review explanations, informing the design of more adaptive and human-centered review tools.
    
    \textbf{RQ\textsubscript{2}.}~\textbf{Which elements of review comments are preferred by developers belonging to different personas?}
    Effective code review comments emphasize different elements, such as explanation depth, conciseness, or learning support, yet it remains unclear whether developers with different problem-solving styles value these elements similarly. Understanding persona-specific preferences is essential for designing effective and inclusive code review feedback. 
    
    \textbf{RQ\textsubscript{3}.}~\textbf{How do developers perceive the usefulness of adaptive, persona-aware code review tools?}
    While persona-aware code review explanations have the potential to better align feedback with developers’ problem-solving preferences, their practical value depends on how developers and reviewers perceive such adaptation. Understanding whether adaptive, persona-aware review tools are viewed as useful, desirable, or concerning is essential before pursuing broader adoption or tool development. 

To address these RQs, we adopt a mixed-methods user study to examine developers' perceptions and assessments of persona-aligned code review explanations. The following sections describe the survey design and the participant recruitment process.

\vspace{-0.3cm}
\subsection{Survey Design}

We use open, multiple-choice, and 5-point Likert scale questions (from 1-strongly disagree to 5-strongly agree). Below, we briefly explain the survey sections:

\textbf{Demographics:} Participants reported their profession, gender, years of development experience, and frequency of participation as code authors and reviewers.

\textbf{Problem-solving style assessment:} In this section, participants were asked how they behave when they approach unknown technology. The questions were used to evaluate participants' GenderMag \cite{burnett2016finding} facets. Based on their responses, participants are mapped to GenderMag personas representing different problem-solving profiles. 
GenderMag is a validated analytical method for identifying inclusivity barriers in software by explicitly considering differences in users’ problem-solving styles. It characterizes variation using three personas, i.e., ``Abi", ``Pat", and ``Tim", across five problem-solving style types: motivation, information processing style, learning style, self-efficacy, and risk attitude. While GenderMag has been applied to evaluate inclusivity issues across various software systems~\cite{murphy2024gendermag, mendez2018gender}, its applicability to developer-facing workflows, such as code review, remains underexplored.
In our study, we focus on the two extreme personas, Abi and Tim, which represent opposite ends of the facet spectrum, and use five representative questions from prior work~\cite{santos2024game} to categorize participants' problem-solving styles based on their responses.

\textbf{Code Snippets:} Participants were presented with three self-contained code snippets, each accompanied by multiple code review comments addressing the same underlying issues but different in style, level of detail, and guidance. The snippets were selected from prior code review datasets \cite{widyasari2025explaining} to represent realistic scenarios and required no project-specific knowledge \cite{guerra2024annotations, dos2018impacts}. We conducted a pilot validation with two developers (with 3 and 7 years of development experience) outside the main study to ensure the snippets were understandable and the functionality explanations were accurate. To further reduce ambiguity, we provided a brief lay explanation of each snippet’s functionality.

\textbf{Code Review Explanation:} To generate persona-aligned code review comments, we used ChatGPT (GPT-5.2) with tailored prompts, following prior work~\cite{santos2025great}. We designed prompts aligned with the GenderMag personas Abi and Tim: Abi-aligned comments provide structured, step-by-step explanations reflecting a process-oriented and risk-averse style, while Tim-aligned comments are concise and action-oriented, encouraging experimentation and autonomy.

Finally, participants were shown the original code review comment along with Abi- and Tim-aligned comments in a randomized order, without any labels indicating their origin. After reviewing each explanation, participants completed a short questionnaire assessing their perceptions, followed by open-ended questions on their preferences, perceived benefits, and concerns regarding persona-aligned code review explanations. After design and proofreading, the survey was piloted with five participants to ensure clarity and understandability. The replication package, including the full survey and prompts used to generate the comments, is available at~\cite{replication}.

\subsection{Participants} 
Participants were recruited via snowball sampling and were required to have prior experience with code review in academic, industrial, or open-source settings. In the first circulation phase, we received 18 non-empty responses, of which 16 passed attention checks and were included in the analysis. The demographics of our participants are summarized in Table~\ref{tab:participant-demographic}. Among the 16 participants, more than half are from academia, with an equal number of novice and expert developers. While 25\% of participants are classified as reviewers, equal proportions are authors or contribute similarly as both authors and reviewers. Finally, participants are evenly split between the Tim and Abi personas.

\subsection{Data Analysis} 
\textit{(I) Likert-scale items:} We measured participants' agreement with code review elements, analyzed the data descriptively and visually. The sample was segmented by problem-solving style (Abi vs Tim), development experience (novice vs expert), and role (author vs reviewer). Following Santos et al.~\cite{santos2021family}, we categorized experience on an ordinal scale. Similar to previous studies that used a 5-year cutoff~\cite{negara2013comparative,wurzel2023competencies}, developers with less than 5 years of experience were classified as novices, and those with 5 or more years as experts. Next, participants were classified as authors or reviewers based on which activity they reported performing more frequently, following the frequency distribution by Ebert et al.~\cite{ebert2021exploratory}. Participants who reported equal involvement in both reviewing and authoring code were assigned to a separate category, labeled \textit{Both}.
\textit{(II) Open questions:} We used open-ended questions to gather deeper insights into participants’ perceptions of adaptive, persona-aware code review tools and qualitatively analyzed their responses.

\section{Preliminary Results} \label{sec:results}

Below, we present results from our ongoing study examining whether personified code review comments can support inclusive code review practices and developers with diverse problem-solving styles.

%
\begin{table}[!ht]
\vspace{-0.3cm}
    \caption{Participant demographics by development experience, profession, role, and persona.}
    \label{tab:participant-demographic}
    \centering
    \vspace{-0.3cm}
        \resizebox{\columnwidth}{!}{
        \begin{tabular}{cc|cc|ccc|cc}
        \toprule
        \multicolumn{2}{c|}{\textbf{Experience}} &
        \multicolumn{2}{c|}{\textbf{Profession}} &
        \multicolumn{3}{c|}{\textbf{Role}} &
        \multicolumn{2}{c}{\textbf{Persona}} \\
        \midrule
        Novice & Expert &
        Dev. & Acad. &
        Author & Reviewer & Both &
        Tim & Abi \\
        \midrule
        8 (50\%) & 8 (50\%) & 7 (44\%) & 9 (56\%) & 6 (38\%) & 4 (25\%) & 6 (38\%) & 8 (50\%) & 8 (50\%) \\
        \bottomrule
        \end{tabular}
    }
    \vspace{-0.3cm}
    \footnotesize{\textit{Note.} Dev. = Developers; Acad. = Academic researcher.}
\end{table}


\textbf{\textit{Preferences for persona-aligned comments (RQ\textsubscript{1}): }}Table~\ref{tab:explanation-preferences} summarizes participants' preferences for persona-aligned code review comments across three code snippets. We observe early indications that explanation preferences vary across personas, experience levels, and code contributors' roles. \textit{Abi} participants selected persona-aligned explanations in roughly half or more of the cases, whereas \textit{Tim} participants showed lower alignment, particularly for the third code snippet. Differences become more pronounced when divided by development experience. 
\textit{Novice-Abi} participants consistently preferred persona-aligned explanations, while \textit{novice-Tim} participants showed limited alignment. 
The higher alignment among \textit{Novice-Abi} participants likely reflects a compound effect of low experience and persona traits. Abi personas have lower self-efficacy and prefer supportive, process-oriented explanations~\cite{santos2023designing}, which can reduce cognitive load for novices and make feedback easier to follow. As a result, \textit{novice-Abi} participants may benefit more from persona-aligned explanations than \textit{novice-Tim} participants.
In contrast, among the expert participants, \textit{Tim} developers exhibited higher alignment than \textit{Abi} developers. 
This result may be explained by Tim's persona, characterized by high confidence in using technology and a belief in being effective at learning new features independently \cite{guizani2022debug}. The lower alignment observed in \textit{Novice-Tim} participants may originate from the lack of technical experience needed to fully benefit from Tim-aligned explanations. Whereas expert participants are more likely to find these explanations useful due to their expertise, a tendency also observed in prior studies on code comprehension as well~\cite{Jessup2021UsingEDA}.
Role-dependent patterns also indicate diverse preference structures, with participants having both author and reviewer roles showing stronger alignment among \textit{Abi} participants. However, the results for \textit{Both-Tim} remain inclusive due to a single participant. Although exploratory and based on a small sample, these findings motivate future large-scale studies and support the vision of an adaptive, persona-aware review system to support diverse developers' needs. 

\begin{table}[!ht]
\centering
\vspace{-0.3cm}
\caption{Percentage of persona-aligned code review explanations chosen across personas, experience levels, and roles.}
\label{tab:explanation-preferences}
    \resizebox{0.8\columnwidth}{!}{
        \begin{tabular}{llccc}
        \toprule
        \textbf{Category} & \textbf{Group} & \textbf{Snippet1} & \textbf{Snippet2} & \textbf{Snippet3} \\
        \midrule
        \textit{Persona} & Tim & 50\% & 50\% & 38\% \\
                & Abi & 50\% & 63\% & 50\% \\
        \midrule
        \textit{Experience} & Novice--Tim & 33\% & 33\% & 33\% \\
                   & Novice--Abi & 60\% & 80\% & 60\% \\
                   & Expert--Tim & 60\% & 60\% & 40\% \\
                   & Expert--Abi & 33\% & 33\% & 33\% \\
        \midrule
        \textit{Role} & Reviewer--Tim & 50\% & 50\% & 25\% \\
             & Reviewer--Abi & 50\% & 50\% & 0\% \\
             & Author--Tim & 0\% & 50\% & 25\% \\
             & Author--Abi & 40\% & 50\% & 0\% \\
             & Both--Tim & 0\% & 0\% & 0\% \\
             & Both--Abi & 40\% & 60\% & 60\% \\
        \bottomrule
        \end{tabular}
    }
    \vspace{-0.3cm}
\end{table}

\begin{center} 
{\setlength{\fboxsep}{6pt}
\colorbox{blue!5!white}{%
  \parbox{0.95\linewidth}{%
    \textbf{Answer to RQ1:} Preferences varied across problem-solving styles (Abi vs. Tim), experiences (novice vs. experienced), and roles (reviewer vs. author vs. both). Abi-aligned explanations were preferred by novice \textit{Abi} participants, while expert \textit{Tim} participants preferred Tim-aligned explanations. 
  }%
}}
\end{center}
\begin{figure}[t]
    \centering
    \begin{subfigure}[t]{\linewidth}
        \centering
        \includegraphics[width=\linewidth]{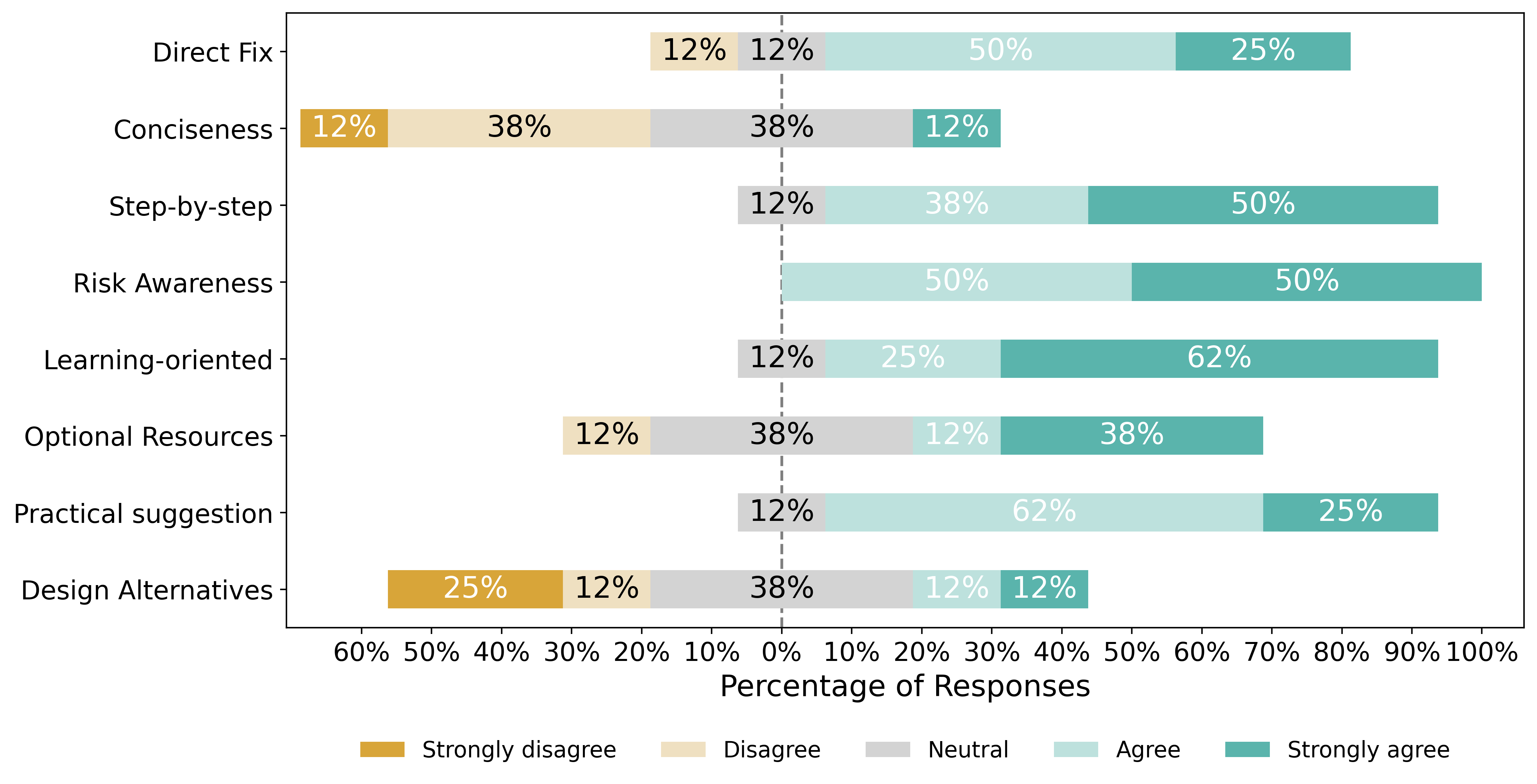}
        \caption{Abi Participants}
        \label{fig:abi-preference}
    \end{subfigure}
    \vspace{0.6em}
    \begin{subfigure}[t]{\linewidth}
        \centering
        \includegraphics[width=\linewidth]{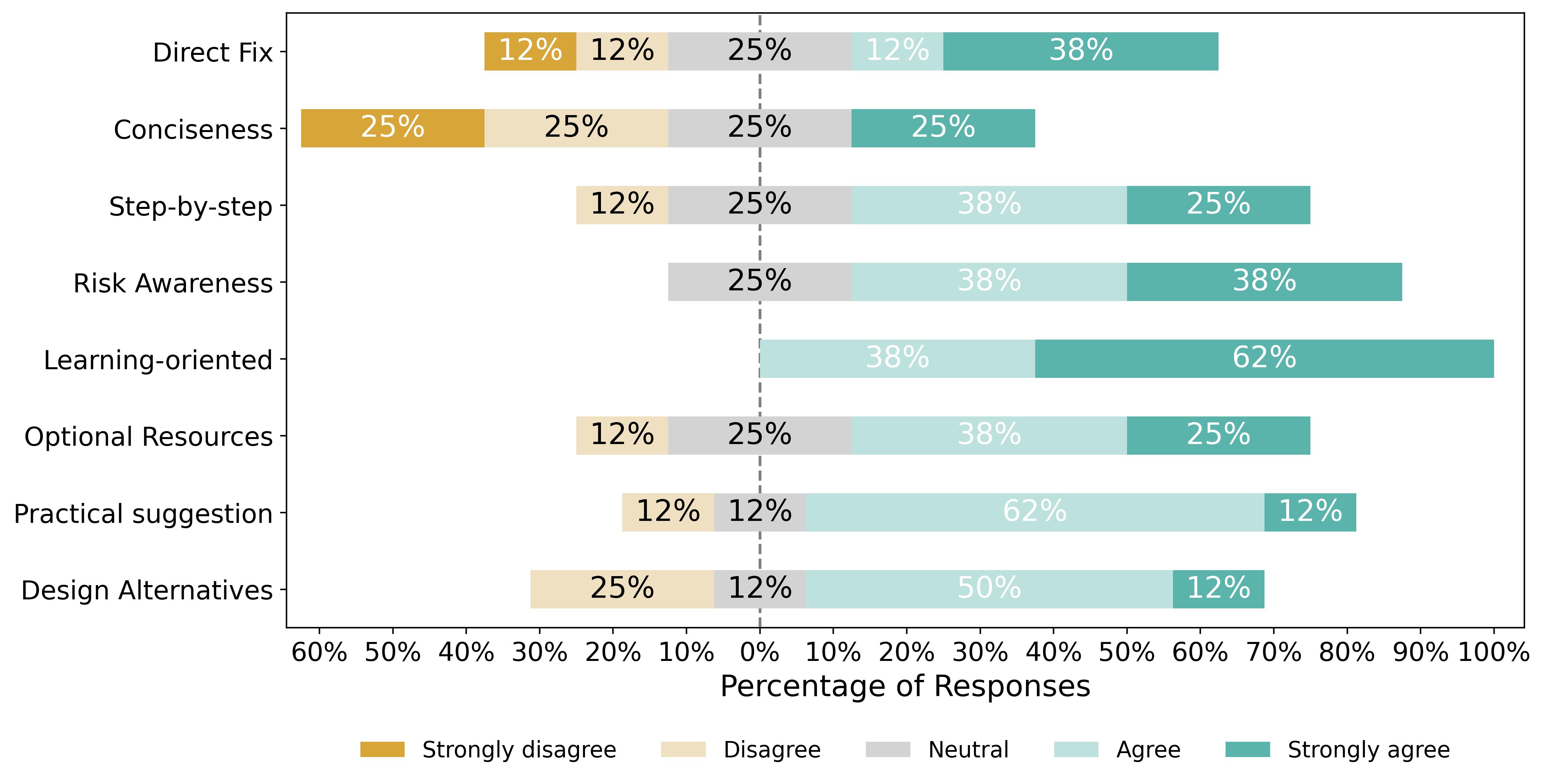}
        \caption{Tim Participants}
        \vspace{-0.5cm}
        \label{fig:tim-preference}
    \end{subfigure}
    \caption{Distribution of participants' preferences for code review comment elements across personas.}
    \label{fig:persona-preference}
\vspace{-0.5cm}
\end{figure}

\textbf{\textit{Preferences in comment elements based on personas (RQ\textsubscript{2}): }} We asked participants to rate their preferred elements in comments. Fig.~\ref{fig:persona-preference} summarizes the participants' agreement with different code review comment elements. Among \textit{Abi} participants, step-by-step explanation, risk awareness, learning-oriented guidance, and practical suggestions received very high agreement, with approximately 80–100\% indicating that they agree or strongly agree. However, concise/brief comments receive largely neutral or negative responses. These preferences, particularly step-by-step explanation, align with Abi persona's process-oriented learning style and lower self-efficacy. \textit{Tim} participants also show similar responses for learning-oriented comments, risk-aware guidance, and practical suggestions, but exhibit more mixed views on conciseness and direct fixes. These early findings indicate that, regardless of persona, developers tend to value explanatory depth, learning support, practical suggestion and risk awareness more than conciseness.
\begin{center} 
{\setlength{\fboxsep}{6pt}
\colorbox{blue!5!white}{%
  \parbox{0.95\linewidth}{%
    \textbf{Answer to RQ2:} Across personas, learning-oriented comments, risk-aware guidance, and practical suggestions were preferred over conciseness, with \textit{Abi} participants additionally valuing step-by-step explanations.
  }%
}}
\end{center}

\textbf{\textit{Perceived usefulness of adaptive code review tools (RQ\textsubscript{3}):}} Most participants expressed a positive view towards a code review tool that adapts comments to individual preferences. Respondents noted that personalization could improve clarity and efficiency by aligning feedback with their problem-solving styles and reducing confusion, especially for novice developers. Several participants highlighted additional perceived benefits, including improved empathy, reduced toxicity, and better support for individuals with specific communication needs, as stated by P8 (an \textit{Abi}), \textit{``Like I said I am autistic so I have very specific communication preferences. So yes, but I am not sure this would help "normal" people.''} At the same time, participants raised concerns about over-simplification, loss of diverse perspectives, and the need for transparency and trust in the adaptation logic. Overall, these responses suggest that adaptive code review tools are promising, but their design must carefully balance personalization with correctness and diversity of feedback.

\begin{center} 
{\setlength{\fboxsep}{6pt}
\colorbox{blue!5!white}{%
  \parbox{0.95\linewidth}{%
    \textbf{Answer to RQ3:} Participants largely viewed adaptive code review tools positively, particularly for novices and developers with specific communication needs, while emphasizing the need to balance personalization with transparency and correctness.
  }%
}}
\end{center}

\section{Related Work} \label{related-work}
Prior work has examined modern code review practices and effective review comments, showing that explanatory and constructive feedback supports developer understanding, while shallow or harsh comments hinder learning~\cite{sarma2024effective, tang2025sphere, widyasari2025explaining}. Other studies categorize review comment intents such as suggestions, defect identification, educational guidance, and design feedback~\cite{chen2025understanding, alami2025engagement, rauf2025human}, but largely assume that such comments are interpreted uniformly by developers. In parallel, research has addressed diversity gaps by modeling problem-solving styles with the GenderMag framework to identify inclusivity issues in software systems~\cite{burnett2021gendermag, santos2023designing, hamid2024measure, santos2024game, culas2025newcomers}. 
More recently, studies have begun exploring LLMs' ability to support developers' diverse learning styles~\cite{brachman2025towards} and have shown that GenderMag can effectively guide LLMs to account for cognitive diversity in designing more inclusive explanations~\cite{santos2025great, anderson2025llm}.

Building on these works, we examine how developers with different problem-solving styles perceive persona-aligned code review explanations, motivating adaptive and inclusive review practices.

\vspace{-0.3cm}
\section{Conclusion and Future Plan} \label{sec:future-plan}
In this paper, we presented a vision for more effective and inclusive code review practices that account for developers’ diverse problem-solving styles. Our preliminary results provided early evidence that developers’ preferences for explanations vary across problem-solving styles, experience levels, and roles.
Our future work will focus on RQ1 and RQ2 to examine persona-aligned explanation preferences across problem-solving styles, while further exploring the feasibility of practical tools (RQ3) to support diverse developer needs. We seek to accommodate the following implications:

\textbf{\textit{Personalized developers support.}} We plan to recruit more participants to investigate further how persona-aligned code review explanations influence developers’ productivity, trust, and decision-making, with more detailed statistical analyses to be reported in a subsequent full paper. This work has the potential to improve accessibility and inclusivity in code review, particularly for developers with specific cognitive or communication needs.

\textbf{\textit{Guiding coding agents toward adaptive reviews.}} Future studies will explore how development experience (novice vs. expert) and review roles (authors vs. reviewers) interact with problem-solving personas. These insights can guide AI-based code review tools and coding agents toward adaptive, persona-aware feedback.

\textbf{\textit{Extending to tool prototyping and feasibility evaluation.}} We plan to design and evaluate a prototype for generating persona-aware code review comments grounded in developers’ observational history and GenderMag personas. We will collect user feedback on the tool’s design and preferred features. The tool will leverage prior interaction patterns, including preferred explanation styles and gaps between provided and preferred comments. We will evaluate usability, perceived usefulness, and workflow integration to assess support for diverse problem-solving styles.

\bibliographystyle{ACM-Reference-Format}
\bibliography{ref}
\end{document}